\documentstyle[11pt]{article}

\newcommand{\be}{\begin{equation}}
\newcommand{\ee}{\end{equation}}
\newcommand{\bea}{\begin{eqnarray*}}
\newcommand{\eea}{\end{eqnarray*}}
\newcommand{\ba}{\begin{eqnarray}}
\newcommand{\ea}{\end{eqnarray}}

\begin{document}

\begin{titlepage}
\begin{flushright}
UCL-IPT-00-17
\end{flushright}

\vspace*{30mm}

\begin{center}
{\Large \bf Trace anomalies and the $\Delta I = {1\over 2}$ rule}
\end{center}

\vspace*{10mm}

\begin{center}
J.-M. G\'erard  and J. Weyers
\end{center}

\vspace*{10mm}

\begin{center}
Institut de Physique Th\'eorique\\
Universit\'e catholique de Louvain\\
B-1348 $\ \ $ Louvain-la-Neuve
\end{center}

\vspace*{50mm}

\begin{abstract}
Trace Anomaly Dominance in weak $K$-decays successfully reproduces
the $\Delta I = {1\over 2}$ selection rule results, as observed in
$K_S \to \pi\pi, K_L \to \pi\pi\pi, K_S \to \gamma\gamma$ and $K_L
\to \pi^0 \gamma\gamma$.
\end{abstract}

\end{titlepage}

\newpage

\section{Introduction}

\hspace*{5mm} A precise quantitative understanding of the $\Delta I
= {1\over 2}$ selection rule in $K$ decays still remains an elusive
goal. Indeed the short-distance evolution of $\Delta S = 1$
dimension six weak operators \cite{1}\cite{2} cannot by itself
reproduce the huge enhancement of the isospin $I=0$ component of
the $K^0 \to \pi\pi$ decay amplitudes relative to the $I=2$ one.

Moreover, pursuing the operator evolution through quantum loops
regularized within the truncated non-linear $\sigma$-model \cite{3}, the
chiral quark model \cite{4} or the extended Nambu and Jona-Lasinio model
\cite{5} introduces large uncontrollable theoretical uncertainties
mainly due to the matching procedure between the (short-distance) Wilson
coefficients and the (long-distance) hadronic matrix elements.

Finally, these phenomenological approaches do not shed light on the
possible contribution of effective hadronic operators which are not
directly accessible through perturbative QCD corrections to the
Fermi current-current Hamiltonian.

As a consequence, the dominant effective Hamiltonian for $\Delta S
=1$ hadronic weak decays remains somewhat heuristic.

In this Letter we advocate a specific non perturbative effect,
namely the {\em trace anomaly}, as the mechanism responsible for
the bulk of the empirical $\Delta I = {1\over 2}$ rule.

We will show that the QCD trace anomaly, already known to dominate
in the $\Psi' \to J/\Psi \pi\pi$ decay \cite{6}, also gives rise to a
large non perturbative $\Delta S = 1$ weak operator which
adequately describes $K \to 2\pi, 3\pi$ decays. The QED
trace anomaly then allows for a parameter free calculation of the
decays $K_S \to 2\gamma$ and $K_L \to \pi\gamma\gamma$ which
agrees with the data.

\newpage

Before concluding this Letter we briefly comment on the
relation and the differences of our somewhat unconventional approach
with more traditional points of view.

\section{Trace anomalies}

\hspace*{5mm} The energy momentum tensor $T_{\mu\nu}$ of a quantum field
theory is most easily obtained \cite{7} from the variation of its action

\be
S = \int d^4 x \sqrt{-g} {\cal L} (x)
\ee
with respect to a non-trivial space-time metric $g^{\mu\nu}$
\be
\delta S \equiv {1\over 2} \int d^4 x \sqrt{-g} T_{\mu\nu} (x) \delta
g^{\mu\nu} (x).
\ee
If we represent the scale transformation
\be
x^\mu \to e^\lambda x^\mu
\ee
as a particular change in the metric
\be
g^{\mu\nu} (x) \to e^{-2\lambda} g^{\mu\nu} (x)
\ee
then the corresponding change in the Lagrangian ${\cal L}(x)$ is
proportional to the trace $T$ of $T_{\mu\nu}$. In other words, scale
invariance requires a traceless energy-momentum tensor.

It is also well-known that the classical scale invariance of gauge
theories with massless matter fields is broken by quantum corrections.
Indeed,  an infinitesimal scale transformation induces a shift in
the renormalized gauge coupling
\be
g(\mu) \to g(e^{-\lambda}\mu) = g(\mu) - \lambda \beta (g)
\ee
such that the corresponding change in the Lagrangian is
\be
{\cal L}(g) \to {\cal L} (g) - \lambda \beta(g){\partial {\cal L}\over
\partial g}.
\ee
Consequently, a trace anomaly arises for the classicaly conserved
dilatation current:
\ba
\partial_\mu (T^{\mu\nu} x_\nu) &\equiv& T^{(m=0)} \nonumber\\
&=& \beta(g) {\partial{\cal L}\over\partial g}.
\ea

For chiral QCD, the trace anomaly reads
\be
T^{(m=0)}_{\mbox{\tiny QCD}} = {\beta(g_s)\over 2g_s} G^a_{\mu\nu}
G^{\mu\nu}_a \ \ \ (a=1,\ldots 8)
\ee
and a similar expression holds for $T^{(m=0)}_{\mbox{\tiny QED}}$ as
well.

Although trace anomalies are intrinsically non perturbative, it is
nevertheless useful for our purposes to give their explicit expression
to lowest order in $\alpha_s$ and in $\alpha$ respectively. For three
flavours of massless quarks this leads to
\ba
T^{(m=0)} &=& T^{(m=0)}_{\mbox{\tiny QCD}} + T^{(m=0)}_{\mbox{\tiny
QED}} \nonumber\\
& \cong& - {9\over 8} {\alpha_s \over \pi} G^a_{\mu\nu} G^{\mu\nu}_a +
{1\over 3} {\alpha\over \pi} F_{\mu\nu} F^{\mu\nu}.
\ea

On the other hand, if we consider the low energy effective lagrangian
for the octet $\pi$ of pseudoscalar Goldstone bosons
\be
{\cal L}_{\mbox{\footnotesize eff}} = {f^2\over 8} \mbox{Tr}
(\partial_\mu U \partial^\mu U^\dagger)
\ee
with
\be
U = \exp i \sqrt{2} {\pi\over f} \ \ \ , \ \ \ f = 132
\ \mbox{MeV}
\ee
the effective strong anomaly is easily computed and the total trace
anomaly is now given by
\ba
T^{(m=0)}_{\mbox{\footnotesize eff}} &=& T^{(m=0)}_{\mbox{\footnotesize
eff, strong}} + T^{(m=0)}_{\mbox{\tiny QED}}\nonumber
\\
&=& - {f^2 \over 4} \mbox{Tr}
(\partial_\mu U \partial^\mu U^\dagger) + T^{(m=0)}_{\mbox{\tiny QED}}.
\ea

\section{The QCD trace anomaly and hadronic $K^0$
decays}

\hspace*{5mm} From Eqs(9) and (12), the gluon conversion into a
(low-energy) system of two pions in a relative $s$-wave is calculable
from first principles and
\be
< \pi^+ \pi^- | T^{(m=0)}_{\mbox{\tiny QCD}} | 0 > = (p_+ + p_-)^2.
\ee

This remarkable property of the strong interactions was elegantly
exploited \cite{6} to predict the two pion invariant mass distribution
in $\Psi' \to (J/\Psi) \pi^+\pi^-$ decays. The new experimental data
from the BES collaboration \cite{8} beautifully confirm this
prediction.

The physical picture underlying this exclusive $\Psi'$ decay is thus a
simple two-step process: emission of soft gluons (in a $0^{++}$ state)
from a {\em heavy} quark and then, via the trace anomaly, hadronization
of these gluons into a pair of pions.

It is of course tempting to invoke a similar mechanism in hadronic $K$
decays, namely emission of soft gluons (in a $0^{++}$ state) by a {\em
light} quark and again hadronization via the trace anomaly. In such a
case, the dominant $\Delta S = 1$ effective Hamiltonian at low energy
would thus read
\be
{\cal H}^{\Delta S = 1}_{\mbox{\tiny TAD}} = g_8 r (M U^\dagger + U
M^\dagger)^{ds}  T^{(m=0)}_{\mbox{\footnotesize eff, strong}}
\ee
where $M =  \mbox{diag} \  (m,m,m_s)$ is the light quark mass matrix
responsible for the pseudoscalar squared masses $m^2_\pi = rm$ and $m^2_K
= {r\over 2} (m+m_s)$, in the isospin limit.

The weak Hamiltonian Eq.(14) has the correct behaviour under both chiral
$SU(3)_L \otimes SU(3)_R$ and CPS \cite{9} transformations. The resulting
$\Delta I = {1\over 2}$ hadronic decay amplitudes are given
by\footnote{Here and in what follows, we always assume CP invariance,
i.e. real $g_8$.}

\ba
A(K_S \to \pi^+ \pi^-) &=& i {4\sqrt{2}\over f} g_8 \left(m^2_K -
m^2_\pi\right) m^2_K  \nonumber \\
A(K_L \to \pi^+\pi^0\pi^-) &=& {4\over f^2} g_8 m^2_K \left({1\over 3}
m^2_K + m^2_\pi Y\right)
\ea
with
$$
Y = {(s_3-s_0)\over m^2_\pi} \ \ \ \ \ , \ \ \ s_0 = {1\over 3}
(s_+ + s_3 + s_-)
$$
the standard Dalitz variables. These amplitudes turn out to be identical
to the ones obtained from the conventional chiral Hamiltonian
\be
{\cal H}^{\Delta  S=1}_\chi = {f^4\over 4} G_8 (\partial_\mu U
\partial^\mu U^\dagger)^{ds}
\ee
provided we make the following identification
\be
G_8 = 4 {m^2_K \over f^2} g_8.
\ee
In particular, we obtain a reasonable (linear) fit of the $K_L \to \pi
\pi\pi$ Dality plot in terms of the $g_8$ parameter extracted from the
measured $K \to \pi\pi$ decay  widths:
\be
|g^{\exp}_8| = 0.16 \ \ 10^{-6} \mbox{Gev}^{-2}.
\ee

\section{The QED trace anomaly and radiative $K^0$
decays}

\hspace*{5mm} If our hypothesis of Trace Anomaly Dominance in the
$\Delta I = {1\over 2} \ K^0$ decays holds true, it is straightforward to
extend Eq.(14) to radiative processes by simply replacing
$T^{(m=0)}_{\mbox{\footnotesize eff, strong}}$ by
$T^{(m=0)}_{\mbox{\footnotesize QED}}$.

The resulting $K_S$ decay amplitude into two real photons is then
\be
A (K_S \to \gamma\gamma) = i {16\sqrt{2}\over 3f} g_8 (m^2_K - m^2_\pi)
{\alpha \over \pi}
\Bigl[(q_1\cdot q_2)
(\varepsilon_1\cdot \varepsilon_2) - (q_1\cdot \varepsilon_2)(q_2\cdot
\varepsilon_1)\Bigr].
\ee
From this equation and Eqs(15), it follows that
\ba
\mbox{Br} (K_S \to \gamma\gamma) &=& \left({2\alpha \over 3
\pi}\right)^2 \left(1-4{m^2_\pi \over m^2_K}\right)^{-{1\over 2}} \
\mbox{Br} (K_S \to \pi^+\pi^-) \nonumber \\
&=& (2.0\pm 0.2) 10^{-6}.
\ea
The quoted error in Eq.(20) corresponds to our neglect of (small)
$\Delta I = {3\over 2}$ contributions.

This result is in good agreement with the recent measurement of NA48
Collaboration \cite{10}
\be
\mbox{Br} (K_S \to \gamma\gamma) = (2.6 \pm 0.5) 10^{-6}.
\ee

In a similar way, we consider the $K_L \to \pi^0 \gamma\gamma$
radiative decay mode. The corresponding amplitude is simply given by
\be
A(K_L \to \pi^0\gamma\gamma) = {16 \over 3f^2} g_8 m^2_K {\alpha \over
\pi} \Bigl[(q_1\cdot q_2)
(\varepsilon_1\cdot \varepsilon_2) - (q_1\cdot \varepsilon_2)(q_2\cdot
\varepsilon_1)\Bigr].
\ee
From Eqs(19), (21) and (22), we obtain the trace anomaly induced
branching
\ba
\mbox{Br} (K_L \to \pi^0\gamma\gamma) &\approx& 0.49 \ \ \mbox{Br} (K_S
\to
\gamma\gamma) \nonumber\\
&=& (1.3 \pm 0.3) 10^{-6}
\ea
in fair agreement with the world average value \cite{11}
\be
\mbox{Br} (K_L \to \pi^0\gamma\gamma) = (1.68 \pm 0.10) 10^{-6}.
\ee

In our approach based on the dominance of trace anomalies, the
$2\gamma$ invariant mass distribution in $K_L \to \pi^0\gamma\gamma$ is
predicted to be negligible at low $z$: with
\be
z \equiv {(q_1+q_2)^2 \over m^2_K}
\ee
the spectrum is indeed given by
\be
{d\Gamma\over dz} = {1\over 36\pi^3f^4} g^2_8 m^9_K z^2
\lambda^{{1\over 2}} \left(1,z,{m^2_\pi\over m^2_K}\right)
\ee
where $\lambda(x,y,z) \equiv x^2+y^2+z^2-2(xy+xz+yz)$ is  the usual
kinematical function\footnote{We use the same notation as in
\cite{12}.}.

\section{Trace Anomaly Dominance versus Chiral
Hamiltonians}

\hspace*{5mm} In Section 3, we already pointed out that the conventional
chiral Hamiltonian ${\cal H}^{\Delta S=1}_\chi$ and the trace anomaly
Hamiltonian
${\cal H}^{\Delta S=1}_{\mbox{\tiny TAD}}$ give identical results for
hadronic
$K$ decays.

The two radiative $K$ decays, on the other hand, are often presented as
significant tests of chiral perturbation theory: pion loops generated
by the chiral Hamiltonian allow these processes to occur.  But for
${\cal H}^{\Delta S=1}_{\mbox{\tiny TAD}}$, two photons are {\em
directly} a piece of the trace anomaly.

For the decay $K_S \to 2\gamma$, the physics of the two pictures is not
that different: the pion loop is indeed an ``effective scalar'' just as
the trace anomaly.

For $K_L \to \pi^0\gamma\gamma$, chiral perturbation theory meets with
some difficulty to account for the measured branching ratio \cite{13} but
not to reproduce the $2 \gamma$ energy spectrum distribution. Once again
the latter fact follows from an effective scalar coupling of the two
photons.

\newpage

Whether the rate problem of $K_L \to \pi^0\gamma\gamma$ is a serious
shortcoming of ${\cal H}_\chi^{\Delta S = 1}$ remains to be seen but, in
any case,
${\cal H}^{\Delta S=1}_{\mbox{\tiny TAD}}$ does agree with all
experimental information presently available.

What, then,  is the dominant effective Hamiltonian for $\Delta S = 1$
weak decays? ${\cal H}^{\Delta S = 1}_\chi$, ${\cal H}^{\Delta S =
1}_{\mbox{\tiny TAD}}$ or a linear combination of both?

The short-distance evolution \cite{1} of the single $\Delta I = {1\over
2}$ four-quark operator
\be
Q_2 -Q_1 \equiv L^{su}_\mu L^\mu_{ud} - L^{sd}_\mu L^\mu_{uu}
\ee
with
$$
L^{ij}_\mu \equiv \bar q^i \gamma_\mu (1-\gamma_5) q^j
$$
down to the charm mass scale obviously favours ${\cal H}^{\Delta
S=1}_\chi$. Indeed, the hadronized left-handed current derived from the
low-energy effective Lagrangian given in Eq.(10) reads
\be
L^{ij}_\mu = {if^2 \over 2} \Bigl(\partial_\mu U U^\dagger\Bigr)^{ji}.
\ee
However, such a hadronization requires first to evolve further down. But
below the charm mass scale, the GIM cancellation mechanism is not
efficient anymore and penguin-like diagrams \cite{2} involving charge
${2\over 3}$ quark loops arise. Among them, an ``annihilation-penguin''
diagram with the {\em heavy} charm quark running inside the loop
induces an effective ${\cal H}^{\Delta S=1}_{\mbox{\tiny TAD}}$. The
resulting perturbative estimate \cite{14} gives a negligible
contribution to the $\Delta I = {1\over 2}$ rule
\ba
g^{\mbox{\tiny SD}}_8 &=& {G_F \over \sqrt{2}} V_{ud} V_{us} {f^2 \over
1080 m^2_c}\nonumber \\
&\approx& 10^{-4} g^{\exp}_8.
\ea
But the {\em same} topology with now a ``soft'' up quark
running in the loop will induce a sizeable enhancement of the $\Delta I
= {1\over 2}$ component of the $K_S \to \pi\pi$ decay amplitude. This is
supported by an estimate based on QCD sum rules \cite{15}.

It seems fair to conclude that we do not have any a priori theoretical
reason for choosing between ${\cal H}^{\Delta S =1}_{\chi}$ and/or
${\cal H}^{\Delta S=1}_{\mbox{\tiny TAD}}$. Of course, a reliable non
perturbative estimate of $G_8$ and $g_8$ would settle the question !

\section{Conclusion}

\hspace*{5mm} We have argued that trace anomalies might in fact dominate
the $\Delta I = {1\over 2}$ component of $K^0$ decay amplitudes. This
rather unconventional approach based on the effective Hamiltonian
defined in Eq.(14) directly predicts $K_S \to \pi\pi$, $K_L \to
\pi\pi\pi$,
$K_S
\to
\gamma\gamma$ and $K_L \to \pi^0\gamma\gamma$ decay amplitudes in terms
of a single parameter $g_8$. All predictions are in good agreement with
the data.

The Trace Anomaly Dominance works beautifully in $\Psi' \to (J/\Psi)
\pi\pi$ decays, and has been suggested \cite{16} as a possible
explanation of the so-called ``$\rho \pi$ puzzle''. For the $\Delta I =
{1\over 2}$ rule in $K$-decays, its contribution cannot be ignored
anymore.

A deeper understanding of the $\Delta I = {1\over 2}$
selection rule appears to be at hand: it depends on (non perturbative)
estimates of our parameter $g_8$ as well as the parameter $G_8$ of the
conventional chiral Hamiltonian. In principle, such an estimate is
accessible to lattice gauge theory.

\end{document}